\documentclass[twocolumn,aps,showpacs,prb,tightenlines,amsmath,amssymb,superscriptaddress]{revtex4}
\usepackage{graphicx}
\usepackage{amssymb}
\usepackage{dcolumn}
\usepackage{amsmath}
\usepackage{bm}
\usepackage{colordvi}

\begin{document}              

\title{Electric manipulation of electron spin relaxation induced by 
confined phonons in nanowire-based double quantum dots}
\author{M. Wang}
\affiliation{Hefei National Laboratory for Physical Sciences at
Microscale, University of Science and Technology of China, Hefei,
Anhui, 230026, China}
\affiliation{Department of Physics,
University of Science and Technology of China, Hefei,
Anhui, 230026, China}
\author{Y. Yin}
\affiliation{Hefei National Laboratory for Physical Sciences at
Microscale, University of Science and Technology of China, Hefei,
Anhui, 230026, China}
\author{M. W. Wu}
\thanks{Author to  whom correspondence should be addressed}
\email{mwwu@ustc.edu.cn.}
\affiliation{Hefei National Laboratory for Physical Sciences at
Microscale, University of Science and Technology of China, Hefei,
Anhui, 230026, China}
\affiliation{Department of Physics,
University of Science and Technology of China, Hefei,
Anhui, 230026, China}
\date{\today}

\begin{abstract}
  We investigate theoretically the electron spin relaxation in
  single-electron nanowire-based semiconductor double quantum dots
  induced by confined phonons and find that the electron spin
  relaxation rate can be efficiently manipulated by external electric
  fields in such system. An anti-crossing, due to the coaction of the
  electric field, the magnetic field and the spin-orbit coupling,
  exists between the lowest two excited states.  Both energies and
  spins of the electron states can be efficiently tuned by the
  electric field around the anti-crossing point.  Multiple sharp peaks
  exist in the electric-field dependence of the spin relaxation rate
  induced by the confined phonons, which can be ascribed to the large
  density of states of the confined phonons at the van Hove
  singularities. This feature suggests that the nanowire-based double
  quantum dots can be used as electric tunable on-and-off spin
  switches, which are more sensitive and flexible than the ones based
  on quantum-well based double quantum dots.  The temperature
  dependence of the spin relaxation rate at the anti-crossing point
  are calculated and a smooth peak, indicating the importance of the
  contribution of the off-diagonal elements of the density matrix to
  the spin relaxation, is observed.
\end{abstract}

\pacs{72.25.Rb, % Spin relaxation and scattering
  73.21.La,     % Quantum dots 
  63.22.-m,     % Phonons or vibrational states in low-dimensional structures and nanoscale materials
  63.20.kd}     % Phonon-electron interactions

\maketitle
\section{INTRODUCTION}

Manipulation of spin relaxation in semiconductor quantum dots (QDs) is
of great interest due to its potential applications in quantum
information processing and spintronic
devices.\cite{jmta,btra,rhans,hanson} Various mechanisms can lead to
the spin relaxation, such as the spin-orbit coupling (SOC) due to the
space inversion asymmetry together with the electron-phonon
scattering,\cite{vngo, lmwo, jlch, khaetskii} the coaction of the
hyperfine interaction and the electron-phonon
scattering,\cite{khaetskii, nazarov, psta,vaab} the g-factor
fluctuations and the direct spin-phonon scattering due to the
phonon-induced strain.\cite{khaetskii} For the widely-investigated
III-V semiconductor QDs, the SOC together with the electron-phonon
scattering dominates the spin relaxation in large magnetic
fields.\cite{vngo,lmwo,jlch,jhji,dvbu} It has been shown that the spin
relaxation can be manipulated not only in single QDs by external
magnetic field,\cite{cfde, ytok} but also in double quantum
dots (DQDs) by external electric field.\cite{jrpe,yywa} The
latter is more interesting since the electric field is more easily
accessible and controllable in genuine  devices. The
electric coherent spin manipulation in DQDs has been demonstrated
experimentally by electric dipole spin resonance with weak electric
field.\cite{nowack, ladriere} It was also suggested that the electron
spin in DQDs can be controlled by strong pumped electric
field.\cite{sherman}

The above mentioned QDs are either self-assembled ones or fabricated
by confining electrons in quantum wells, where bulk phonons in
substrates play an important role.\cite{hanson, vngo,jhji} In recent
years, a new type of QDs based on self-assembled nanowires has been
fabricated.\cite{bjork,bjork2,nilsson,mtbj,cfas} The nanowires are
perpendicular to the substrates, making the bulk phonons in the
substrates less important than the quasi-one-dimensional confined
phonons in the nanowires.\cite{hsht,bjoh} It has been shown that the
confined phonons can lead to novel properties in optical absorption in
nanowire-based single QDs and transport through nanowire-based
DQDs.\cite{cgal,glin,cweber1} Specifically, the spin relaxation in
nanowire-based single QDs can be efficiently manipulated by external
magnetic field due to the large phonon density of states (DOS)
near the van Hove singularity of the confined phonons.\cite{yyin}
However, the manipulation of spin relaxation by electric field
in these nanowire-based QDs is not reported yet.

In this paper, we propose a scheme of manipulation of the spin
relaxation in nanowire-based elongate DQDs by external electric
field. In our system, the electric field can efficiently tune the
energy spectrum of the DQDs where an anti-crossing between electron
states with opposite majority spin exists. As the energy splittings
between these states match certain van Hove singularity of the
confined phonons, multiple sharp peaks occur in the electric field
dependence of the spin relaxation rate (SRR). This feature offers an
efficient scheme for the manipulation of the spin relaxation and hence
on-and-off switches. Comparing to the electric spin relaxation
manipulation in quantum-well-based DQDs,\cite{yywa} the SRR in our
system is much more sensitive to the electric field and can be
manipulated with more flexibility.  Moreover, due to the strong spin
mixing in the vicinity of the anti-crossing point, the off-diagonal
density matrix elements have a large contribution to the SRR. Such
contribution can have a pronounced impact on the SRR, which can
  be seen in the temperature dependence of the SRR. It should
be noted that the strong spin mixing makes it difficult to define the
majority spin for each state in the vicinity of the anti-crossing
  point. This makes the widely-used Fermi-golden-rule
approach\cite{lmwo, khaetskii, nazarov, jlch,jhji,psta,dvbu,cfde,vaab}
for spin relaxation calculation become obscure in this regime. So in
the calculation, we apply the equation-of-motion approach developed by
Jiang {\em et al.}.\cite{jhji}

We organize the paper as follows. In Sec.~II, we introduce the model
for our system and specify the Hamiltonian. The equation-of-motion
approach is also briefly described in this section. In Sec.~III, we
first present the electric field dependence of the energy levels and
the corresponding spin states of the DQDs. Then we show the electric
field dependence of the SRR and demonstrate the electric manipulation
of spin relaxation. Finally, we discuss the impact of the
off-diagonal elements of the density matrix
on the temperature dependence of the SRR. The validation of the
Fermi-golden-rule is also addressed in this section. We summarize in
Sec.~IV.

\section{MODEL AND FORMALISM}

We consider a single-electron elongate DQDs embedded in InAs cylindrical
nanowire with radius $R$. An external magnetic field $\bm{B}$ is applied along
the wire (the $z$-axis). The total Hamiltonian can be expressed as
\begin{eqnarray} 
  H_{\rm tot} & = & H_{\rm e} + H_{\rm ph} + H_{\rm ep},
  \label{eq1}
\end{eqnarray} 
where $H_{\rm e}$, $H_{\rm ph}$ and $H_{\rm ep}$ are the
electron, phonon and electron-phonon interaction
Hamiltonian, respectively. The electron
Hamiltonian has the form $H_e=H_0+H_{\rm so}$. Here $H_0$ is electron
Hamiltonian without the SOC:
\begin{eqnarray} 
  H_{0}&=&\frac{{\bf{p}}^2}{2m^{\ast}}+V_{c}(r)+V_{z}(z)+H_{\rm B},
\end{eqnarray} 
with $m^{\ast}$ denoting the electron effective mass. $V_{\rm
  c}(r)=\frac{1}{2}m^{\ast}\omega_{0}^2 r^2$ is the QD confinement in radial
direction. Thus the dot has an effective diameter $d_{0}=\sqrt{\hbar
  \pi/m^{\ast}\omega_{0}}$. The confinement $V_{z}(z)$ in 
the $z$-axis is given by
\begin{equation} 
  V_{z}(z)= 
  \begin{cases}
    eEz+\frac{1}{2}e(a+2d)E, & \frac{1}{2}a<|z|<\frac{1}{2}a+d, \\
    eEz+\frac{1}{2}e(a+2d)E+V_{0}, & |z|\le \frac{1}{2}a, \\ 
    \infty, &\text {otherwise},
  \end {cases}
\end {equation} 
with $V_{0}$ representing the barrier height between the two dots, $a$ standing
for the interdot distance and $d$ denoting the dot length. A schematic of
$V_{z}(z)$ is plotted in Fig.~\ref{fig1}(b)(IV). In our model, the radial
confinement is much stronger than the longitudinal one, so we consider only the
lowest electron subband in the radial direction. $H_{B}=\frac{1}{2}g\mu_B \bm{B}
\cdot \bm{\sigma}$ is the Zeeman term with $g$, $\mu_B$ and $\bm{\sigma}$
representing the electron g-factor, Bohr magneton and Pauli matrix,
respectively. The SOC Hamiltonian takes the form
\begin{eqnarray}
  H_{\rm so} & = & \frac{\gamma}{\hbar}\sigma_yp_z, 
\end{eqnarray}
where $\gamma$ is the Rashba strength.\cite{mirc,clro}

The confined phonon modes are calculated within isotropic elastic continuum
model by assuming free-surface boundary
conditions.\cite{ancl,hsuz,pcha,syuk,smko,ttak,baau,mast} The Hamiltonian is
given by $ H_{\rm ph}=\sum_{q \lambda}\hbar\omega_{q \lambda}a^{\dagger}_{q
  \lambda}a_{q \lambda}$, where $\omega_{q \lambda}$ is the eigenfrequency of
the $\lambda$th phonon mode with axial wave vector $q$, which is solely
  determined by the longitudinal or transverse sound velocity $v_L$ 
or $v_T$, respectively. Note that the calculated confined phonon modes can be
  classified into axial and radial modes according to the corresponding
  displacement field at small $q$.\cite{cweber, cweber1}

Both the deformation potential and piezoelectric couplings are
considered in our calculation. Assuming the displacement field of the confined
phonons is $\bm{u}(\bm{r})$, the deformation potential coupling is given by the
divergence of $\bm{u}(\bm{r})$: $H_{\rm ep}^{\rm D} (\bm{r})= - \Xi{\bm
  {\nabla}}\cdot {\bm u}(\bm{r})$, where $\Xi$ is the deformation coupling
strength. The piezoelectric coupling is determined via $H_{\rm ep}^{\rm
  P}(\bm{r})=\frac{e}{\kappa} \int d{\bm r}_{e} \frac{{\bm \nabla} \cdot {\bm
    P}^{\rm pz}}{\left| {\bm r}-{\bm r}_{e} \right|}$, where $\kappa$ denotes
the relative dielectric constant, $e$ is the electron charge and ${\bm P}^{\rm
  pz}$ stands for the polarization induced by $\bm{u}(\bm{r})$.\cite{ttak} For
InAs nanowires with wurtzite structure, ${\bm P}^{\rm pz}({\bm
  r})=e_{\rm 15}[\partial_{z}u_{r}({\bm r})+\partial_{r}u_{z}({\bm r})]{\bm
  e}_{r}+[e_{\rm 31}(\partial_{r}+\frac{1}{r})u_{r}({\bm
  r})+e_{\rm 33}\partial_{z}u_{z}({\bm r})]{\bm e}_{z}$, where $e_{\rm 15}$,
$e_{\rm 31}$
and $e_{\rm 33}$ are the piezoelectric constants.\cite{cweber, adby} By
  substituting $\bm{u}(\bm{r})$ into the expression of $H_{\rm ep}^{\rm D}$ and
  $H_{\rm ep}^{\rm P}$,\cite{cweber} one can express the Hamiltonian of the
  electron-phonon interaction in the form
\begin{equation} 
  H_{\rm ep}=\sum_{\eta=D,P} \sum_{{q}\lambda}M_{ q \lambda}^{\eta}\chi_{{q}\lambda}^{\eta} (a_{{q}\lambda}+
  a^{\dagger}_{{-q}\lambda}),
\end{equation} 
where $|M^{\rm D}_{q\lambda}|^2=\hbar\Xi^2
  /(2\pi\rho_m\omega_{q\lambda}R^2)$ for the deformation potential coupling and
  $|M^{\rm P}_{q\lambda}|^2=8\hbar\pi e^2e^2_{\rm 14
  }/(\kappa^2R^2\omega_{q\lambda})$ for the piezoelectric coupling, with
  quantity $\rho_m$ being the mass density. The factor
$\chi_{{q}\lambda}^{\eta}(\bm{r})$ depends on the confined phonon
eigenmode. Note that since in the calculation, we consider only the lowest
radial subband of the electron, only the dilatation phonon modes couple to the
electron.\cite{cweber1,nnis}

Given the electron Hamiltonian $H_e$, the eigenstates $\{|\ell\rangle\}$ and
corresponding eigenenergies $\{\varepsilon_{\ell}\}$ can be obtained by
exact-diagonalization method.\cite{jlch} The spin of each state can be
calculated as $\langle \ell | \bm{S} | \ell \rangle$ with
$\bm{S}=\frac{1}{2} \bm{\sigma}$. We apply the equation-of-motion approach to
calculate the spin relaxation time. Following Ref.~\onlinecite{jhji}, we use the
density matrix $\rho=\rho^{e}\otimes\rho^{\rm ph}$ to describe the system. Its
time evolution in the interaction picture reads
\begin{eqnarray}
  \partial_{t}{\rho}^{\rm I}(t)& =& -\frac{i}{\hbar}[V^{\rm
    I}(t),{\rho}(0)]\nonumber \\ 
  &&\mbox{} -\frac{1}{\hbar^2} \int^t_0d\tau[V^{\rm I}(t),[V^{\rm I}(\tau),{\rho}^{\rm I}(\tau)]].
\end{eqnarray} 
By applying the Born-Markovian approximation and carrying out the integral over
the constant energy surface of the confined phonons, the equation-of-motion of
the density matrix $\rho^e={\rm tr_{ph}}(\rho)$ for electron can be obtained from the above equation
within the basis of the electron eigenstates (in Schr\"odinger
picture)\cite{jhji}
\begin{eqnarray}
  \partial_{t}\rho^{\rm e}_{ mn} & = &
  -i\frac{\varepsilon_m-\varepsilon_n}{\hbar} \rho^{\rm e}_{
    mn}-\frac{1}{2}\sum_{\eta=D,P}\sum_{kl}\Big\{W_{mkkl}^{\eta}(\varepsilon_{lk})
  \rho^{\rm e}_{ln}\nonumber\\
  &&\mbox{} -W_{mkln}^{\eta}(\varepsilon_{ln})\rho^{\rm e}_{kl}+ {\rm H.c.}\Big\},
\end{eqnarray}
with
\begin{eqnarray}
W_{mkln}^{\eta}(\varepsilon_{ln})&=&\sum_{q\lambda}\frac{|M^{\eta}_{q\lambda}|^2\chi_{q\lambda}^{\eta
      mk}\chi_{-q\lambda}^{\eta ln}}{\big| \partial_q\omega_{q\lambda}
    \big|} \nonumber\\
&&\mbox{}\times \Big\{ [\bar{n}
  (\omega_{{q}\lambda})+1]\theta(\varepsilon_{nl})\Big
  |_{\hbar\omega_{q\lambda}=\varepsilon_{nl}} \nonumber\\
&&\mbox{} +\bar{n}(\omega_{q \lambda}) \theta ( \varepsilon_{ln} )
  \Big|_{\hbar\omega_{q\lambda}=\varepsilon_{ln}} \Big\},
  \label{eq10}
\end{eqnarray}
in which $\chi^{\eta mn}_{q\lambda}=\langle m|\chi^{\eta}_{q\lambda} |n
\rangle$ and $\theta(\varepsilon_{ln})$ denotes the step function with
$\varepsilon_{ln}=\varepsilon_l-\varepsilon_n$. Note that the transition
  rate $\Gamma_{\ell_1\ell_2}^{\eta}$ from state $|\ell_1 \rangle$ to $|\ell_2
  \rangle$ can be expressed as
  $\Gamma_{\ell_1\ell_2}^{\eta}=W_{\ell_2\ell_1\ell_1\ell_2}^{\eta}
(\varepsilon_{\ell_2\ell_1})$.

By applying a proper initial state, one can use the equation-of-motion of the
density matrix to calculate the time evolution of $\langle S_z
\rangle=\rm{tr}[\hat S_z\rho^e(t)]$, which is the expectation value of the
$z$-component of the electron spin. We assume that the initial state is prepared
by electrical spin injection\cite{ohno, loffler} or optical pumping of the
electron spin,\cite{atature, press, shabaev} which can polarize the electron
spin along the external magnetic field. Thus density matrix elements in the
basis of $\{|\ell\rangle\}$ for the initial state can be expressed as
$\rho^e_{\ell \ell'}(0)=\sum_{\xi} \langle \ell | \xi \rangle F_{\xi} \langle
\xi| \ell' \rangle$, where $\{| \xi \rangle\}$ are the spin-down eigenstates
($\langle \xi | S_z | \xi \rangle=-1/2$) of $H_0$ with eigenenergies
$\{\varepsilon_{\xi}\}$.  $F_{\xi}=C {\rm exp}[-\varepsilon_{\xi}/(k_BT)]$ is
the Maxwell-Boltzmann distribution with $C$ denoting the normalization
parameter. Note that due to the SOC, $\rho^e(0)$ can have off-diagonal elements.

Given the time evolution of $\langle S_z \rangle$, we define the effective
  SRR as 
\begin{equation}
\bar{\tau}^{-1}_{\rm \alpha}=({\rm ln}\alpha)/\tau_{\alpha}
\end{equation}
 with  $\tau_{\alpha}$ being the time that the envelope of $\langle S_z \rangle$
  decays to the $1/\alpha$ from its initial value to its equilibrium value. Note
that comparing to the SRR used in previous works,\cite{jhji, opt_ori} we
introduce an additional factor ${\rm ln}\alpha$ in order to guarantee
that if $\langle S_z \rangle$ can be well-described by a single exponential
decay, $\bar{\tau}^{-1}_{\rm \alpha}$ is independent on $\alpha$. This is
 helpful when we compare the effective SRR $\bar{\tau}^{-1}_{\rm \alpha}$ with
 different $\alpha$. It should be emphasized that if $\langle S_z \rangle$ does
 not follow a single exponential decay, the effective SRR $\bar{\tau}^{-1}_{\rm
  \alpha}$ is $\alpha$-dependent. One should properly choose $\alpha$ 
to give an overall description of the spin relaxation in the regime
  of investigation.

\section{NUMERICAL RESULTS}
In the computation, we consider a DQD with length $d=50$~nm for each dot,
interdot distance $a=10$~nm, barrier height $V_0=0.3$~eV and effective dot
diameter $d_0=12$~nm. The Rashba strength $\gamma$ is set to be $2.4\times
10^{-11}$~eV$\cdot$m, corresponding to the spin-orbit length $\lambda_{\rm
  so}=\hbar^2/(m^{\ast}\gamma)=133$~nm which is close to the value
$\lambda_{\rm{so}} \sim 127$~nm reported in Ref.~\onlinecite{cfas}. The radius
of nanowire $R$ is set to be $30$~nm. The InAs nanowires have wurtzite crystal
structure with orientation along the [0001] direction.\cite{cfas,apfu} The
piezoelectric constants for the wurtzite InAs can be expressed as $e_{\rm
  15}=e_{\rm 31}=-e_{\rm 33}/2=-e_{\rm 14}/\sqrt{3}$ with $e_{\rm 14}$ being the
piezoelectric constant for bulk InAs.\cite{cweber} Note that although both the
deformation potential and piezoelectric couplings are considered in our
computation, we find that the piezoelectric coupling is dominant in our
system. The other parameters used in our calculation
are listed in Table I.

\begin{center}
  \setlength{\unitlength}{1mm}
  \begin{picture}(85,45) \put(1,37){\shortstack[l]{TABLE I. Material parameters
        used in the computation \\(from Ref.~\onlinecite{semi} unless otherwise
        specified).}}  \thinlines 
    \put(0,6){\line(60,0){85}}
    \put(0,6.6){\line(1,0){85}} 
    \put(0,34){\line(1,0){85}}
    \put(0,34.6){\line(1,0){85}} 
    \put(2,9){\shortstack[l]{$\rho_m$\\\\\\\\
        $\upsilon_T$ \\\\\\\\\\$\upsilon_L$ \\\\\\\\\\$e_{\rm
          14}$}}
    \put(16,9){\shortstack[l]{$5900$~kg/m$^3$\\\\\\$2130$~m/s\\\\\\$4410$~m/s
        \\\\\\$3.5 \times 10^8$~V/m}}
    \put(50,10){\shortstack[l]{$\Xi$\\\\\\$m^{\ast}$\\\\\\\\\\$\kappa$\\\\\\\\\\$g$}}
    \put(65,9.){\shortstack[l]{$5.8$~eV\\\\\\\\$0.0239$~$m_0$\\\\\\\\$15.15$\\\\\\\\$-9.0$$^{\rm a}$}}
%    \put(0,6){\shortstack[l]{$^{\rm a}$ $m_0$ is the free electron mass.}}  
    \put(0,1){\shortstack[l]{$^{\rm a}$ Ref.~\onlinecite{cfas}.}}
  \end{picture}
  \vspace{0.3cm}
\end{center}

\subsection{Energy levels and spin configurations}

\begin{figure}
  \centering
  \includegraphics[width=7.5cm]{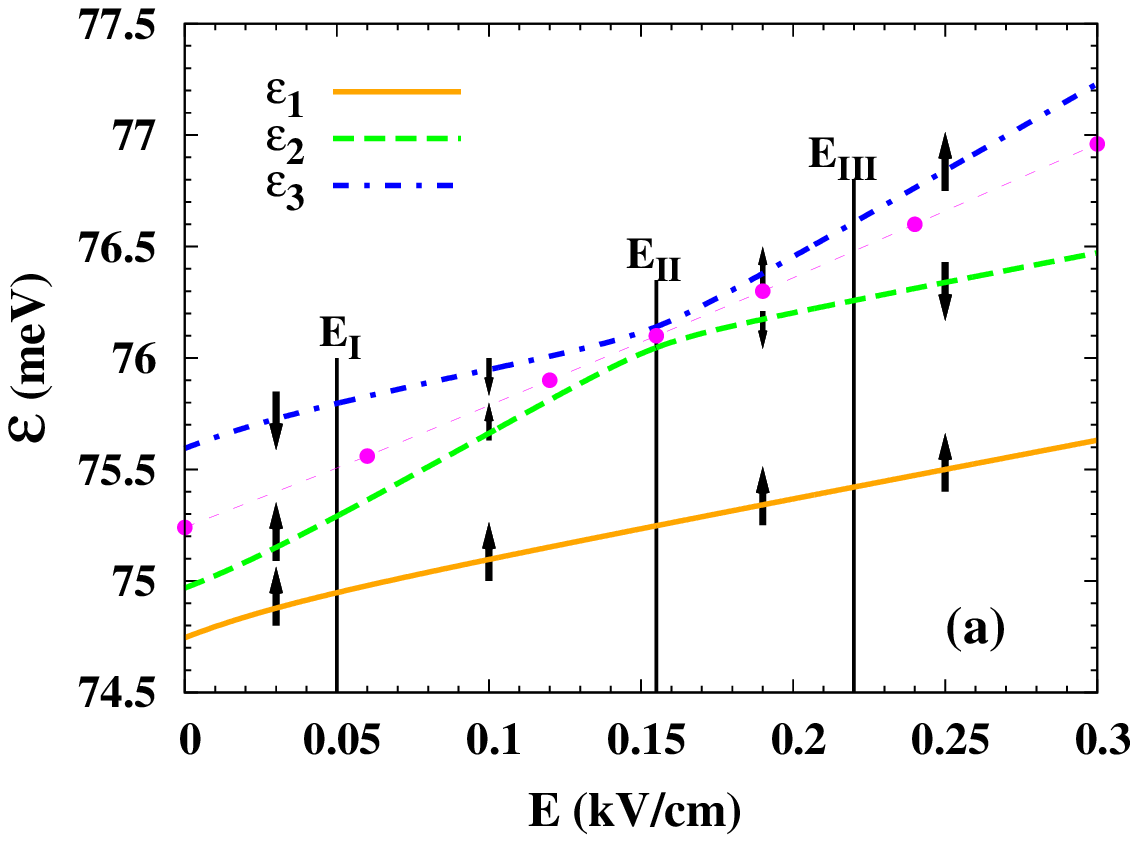}
  \includegraphics[width=7.5cm]{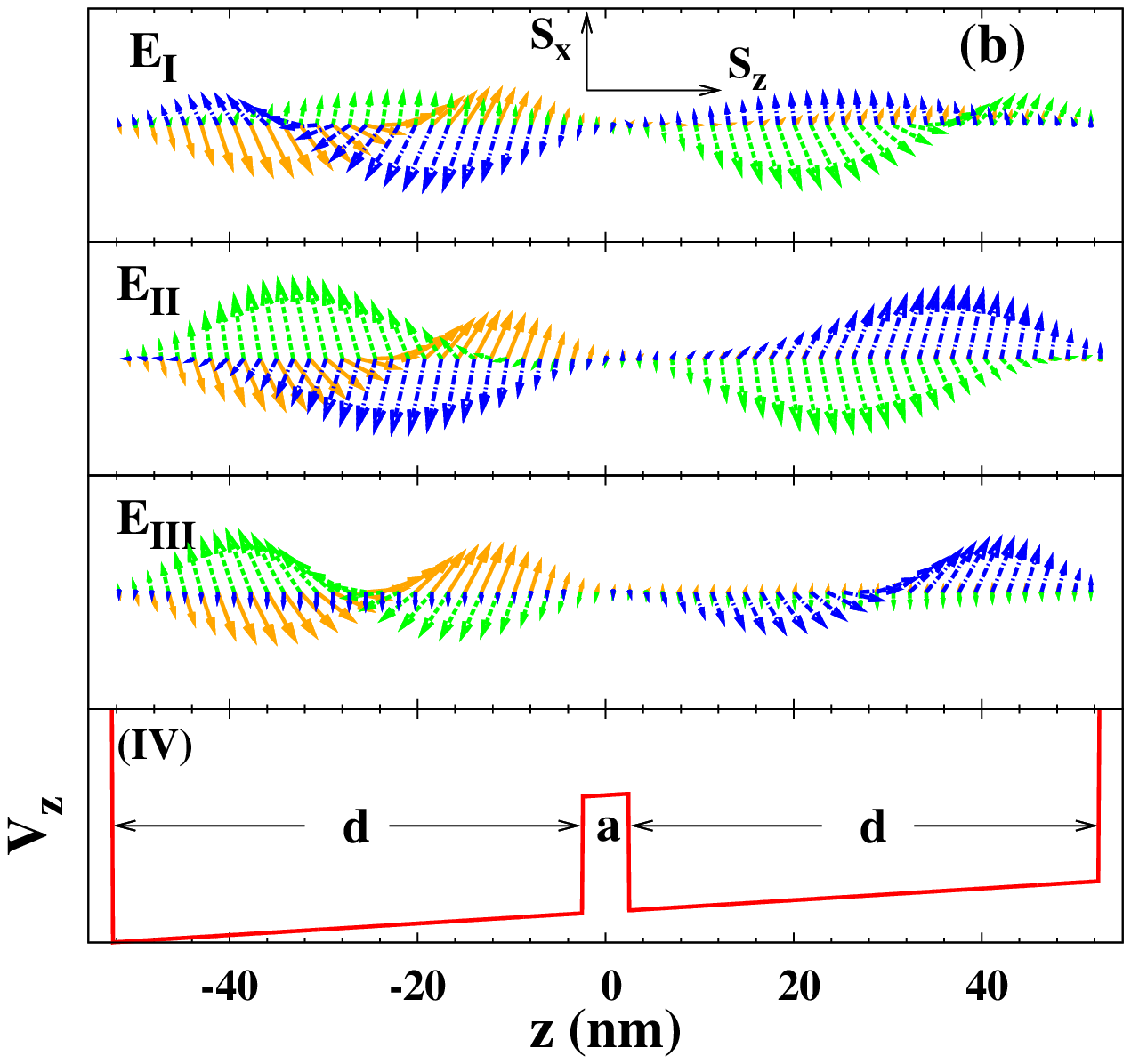}
  \caption{(Color online) (a) The lowest three energy levels vs the electric
    field $E$ with the magnetic field $B=1$~T. The arrows indicate the spins of
    the states. Longer arrow means larger spin. The thin pink dashed curve with
    dots represents the position of the anti-crossing point between the states
    $|2\rangle$ and $|3\rangle$ at different magnetic fields, where the
    corresponding magnetic fields of the dots are $0.27$~T, $0.46$~T, $0.81$~T,
    $1.0$~T, $1.16$~T, $1.53$~T and $1.9$~T, respectively (from left to
    right). The three vertical lines label the typical electric fields $E_{\rm
      I} = 0.05$~kV/cm, $E_{\rm II}=0.156$~kV/cm and $E_{\rm III}=0.22$~kV/cm
    corresponding to the regime before, at and after the anti-crossing
    point. (b) The spin configurations of the lowest three energy levels as
    function of $z$ for the three typical electric fields $E_{\rm I}$, $E_{\rm
      II}$ and $E_{\rm III}$. Arrow in (b) and curve in (a) with the same color
    and line type stand for the same state. The schematic of the potential of
    the DQD is shown in (b)(IV).}
  \label{fig1}
\end{figure}

Before we study the behavior of the effective SRR, it is helpful to
first discuss the electric field dependence of energies and spins of
electron states in the DQD. We plot the lowest three energy levels of
the DQD as function of electric field in Fig.~\ref{fig1}(a) with
external magnetic field $B=1$~T. An anti-crossing between the first
excited state $|2\rangle$ and the second excited state $|3\rangle$ can
be clearly identified in the energy levels which is induced by the
coaction of the electric field, the magnetic field and the SOC. The
position of the anti-crossing point depends on the magnetic field,
which is shown in Fig.~\ref{fig1}(a) by dots at some selected magnetic
fields. The arrows in Fig.~\ref{fig1}(a) illustrate the spin of each
state along the $z$-axis defined as $\langle \ell | S_z | \ell
\rangle$ (The other spin components are zero). With the increase of
the electric field, the spins of the lowest three levels ($|1\rangle$,
$|2\rangle$ and $|3\rangle$) change from $\uparrow$, $\uparrow$ and
$\downarrow$ (before the anti-crossing point) to $\uparrow$,
$\downarrow$ and $\uparrow$ (after the anti-crossing point),
respectively. In the vicinity of the anti-crossing point, due to the
spin mixing induced by the SOC, the spins of the states $|2\rangle$
and $|3\rangle$ decrease and tend to zero at the anti-crossing point,
while the spin of $|1\rangle$ is not sensitive to the electric field.

The electron spin states can be better illustrated by the spin configuration of
each state, e.g., the distribution of each spin state as function of $z$. We
plot the spin configurations of the three states in Fig.~\ref{fig1}(b) at three
typical electric fields $E_{\rm I}$/$E_{\rm II}$/$E_{\rm III}$ corresponding to
the fields before/at/after the anti-crossing point. The component $\langle
S_{y}\rangle$ of the spin configuration is always zero for all the three states
so it is not plotted in the figure. One finds that the spin configuration of the
ground state is not sensitive to the electric field and it is mainly distributed
in the left dot for not too small electric field ($E > 0.01$~kV/cm). The effect
of the electric field on the spin configuration is pronounced for the two
excited states. There are two types of spin configurations for the excited
states far away from the anti-crossing point: (i) the spin configuration
distributes mainly in the right dot and has a similar structure as the spin
configuration of the ground state in the left dot; (ii) the spin configuration
distributes mainly in the left dot and has a different structure as the ground
state. Before the anti-crossing point [Fig.~\ref{fig1}(b)(E$_{\rm I}$)],
$|2\rangle$/$|3\rangle$ belongs to type (i)/(ii). In contrast, after the
anti-crossing point [Fig.~\ref{fig1}(b)(E$_{\rm III}$)], $|2\rangle$/$|3\rangle$
belongs to type (ii)/(i). In the vicinity of the anti-crossing point
[Fig.~\ref{fig1}(b)(E$_{\rm II}$)], both excited states belong to a new type of
spin configuration, which is evenly distributed in the DQD and polarized mainly
along the $x$-axis. This suggests that there is a strong spin mixing between the
two excited states, which makes the expectation value of the electron spin
$\langle S_z \rangle$ associate with not only the diagonal elements but also the
off-diagonal elements between the two states in the density matrix.

As both the energies and spins of the two excited states can be efficiently
manipulated in the vicinity of the anti-crossing point, one would expect that
they should have a pronounced impact on the spin relaxation. So in the following
discussion, we will concentrate on the lowest three levels.

\subsection{Effective spin relaxation rate}

With the equation-of-motion approach we applied here, the effective SRR is
obtained from the time evolution of $\langle S_z \rangle$, which has different
behaviors under different electric/magnetic fields and temperatures. So we start
our discussion of the spin relaxation by studying the typical time evolution of
$\langle S_z \rangle$. Then we study the electric dependence of the effective
SRR and demonstrate how the electric spin relaxation manipulation can be
achieved in our system. We also compare our scheme to the scheme based on DQDs
in quantum wells.  Finally, by studying the temperature dependence, we also show
the importance of the off-diagonal elements of the density matrix to the spin
relaxation.

\subsubsection{Typical behavior of $\langle S_z\rangle$}

\begin{figure}
  \begin{center}
    \includegraphics[width=7.5cm]{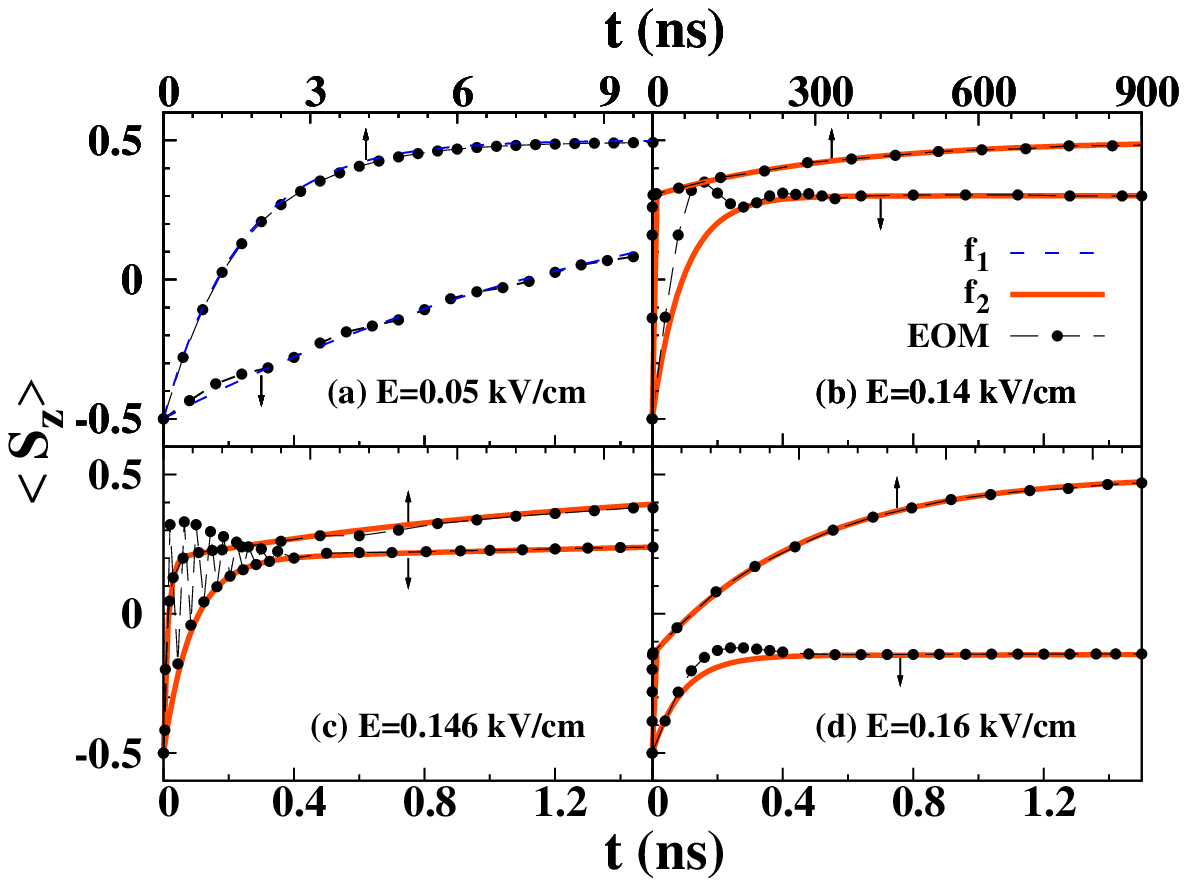}
    \includegraphics[width=7.5cm]{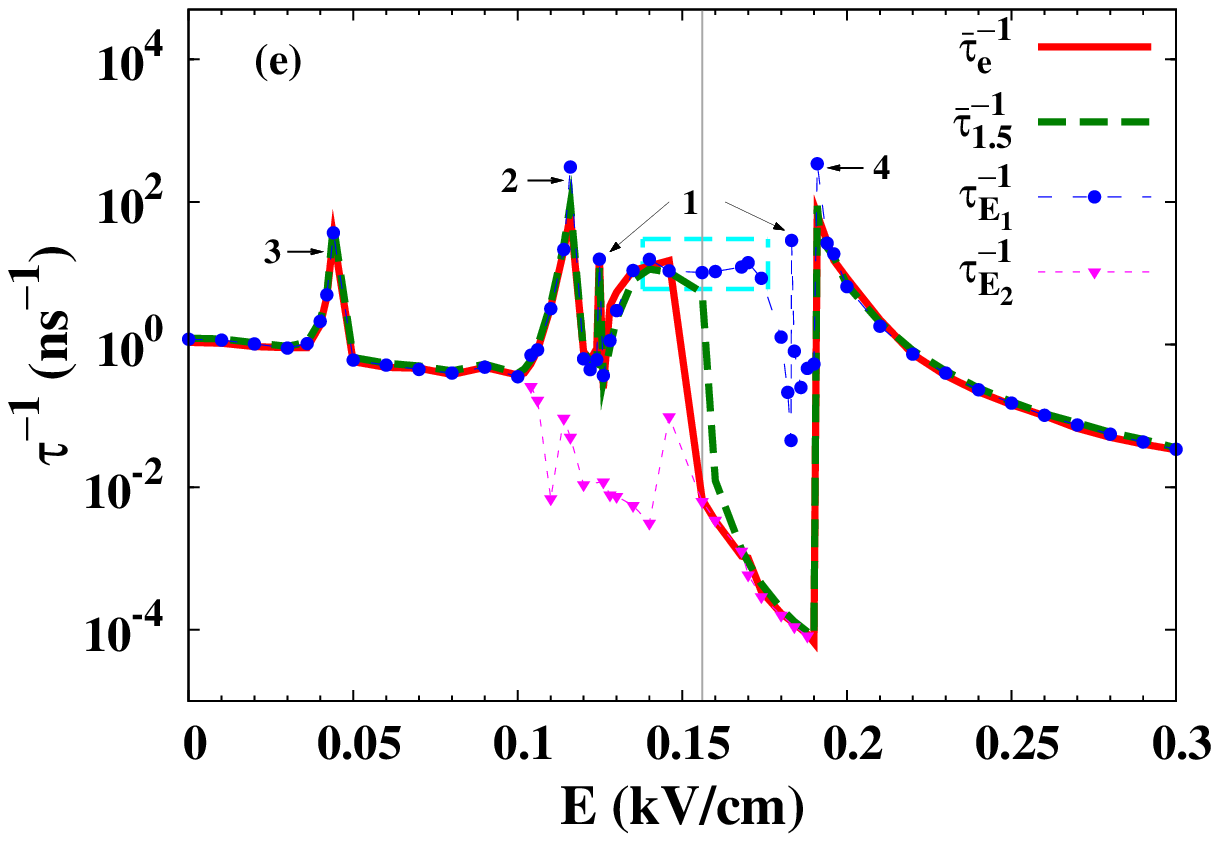}
  \end{center}
  \caption{(Color online) (a) $\sim$ (d) Time evolution of $\langle S_z \rangle$
    in both the short- (lower scale) and long-time (upper scale) regimes with
    $B=1$~T and $T=0$~K at the electric field $E=0.05$, 0.14, 0.146 and
    0.16~kV/cm, respectively. Black dashed curve with dots: $\langle S_z
    \rangle$ from the equation-of-motion approach. Blue dashed curve:
    best-fitting with single exponential decay. Red solid curve: best-fitting
    with double exponential decay. (e) SRRs as function of electric field with
    $B=1$~T and $T=0$~K. Red solid/Green dashed curve: effective SRR
    $\bar{\tau}^{-1}_{\rm{e}}$/$\bar{\tau}^{-1}_{\rm{1.5}}$. Blue
    dots/Pink triangles: rate for the first/second spin relaxation process. The
    cyan dashed box indicates the vicinity of the anti-crossing point. The gray
    vertical line indicates the position of the anti-crossing point.}
  \label{fig2}
\end{figure}

We show the typical behaviors of time evolution of $\langle S_z \rangle$ in both
the short- and long-time regimes at different electric fields in
Fig.~\ref{fig2}(a)$\sim$(d). Generally speaking, they can be classified into two
types: (i) single-exponential decay [Fig.~\ref{fig2}(a)] and (ii)
double-exponential decay [Fig.~\ref{fig2}(b), (c) and (d)]. For type (i), the
time evolution of $\langle S_z \rangle$ [dots in Fig.~\ref{fig2}(a)] can be
well-fitted by a single-exponential function $f_1(t)=a_1 e^{-t/
  \tau_{\rm{E_1}}}+b_1$ [blue dashed curve in Fig.~\ref{fig2}(a)] with $a_1$,
$b_1$ and $\tau_{\rm{E_1}}$ being fitting parameters. For type (ii), the time
evolution of $\langle S_z \rangle$ [dots in Fig.~\ref{fig2}~(b), (c) and (d)]
has two stages: a short-time stage where $\langle S_z \rangle$ decays faster and
a long-time stage where $\langle S_z \rangle$ decays slower. The relative
importance of the two stages can be different. For example, the short-time stage
is important in Fig.~\ref{fig2}(b) since it leads to a large variation of
$\langle S_z \rangle$ from $-0.5$ to $0.3$. While in Fig.~\ref{fig2}(d), the
long-time stage is important. Note that usually the time evolution of $\langle
S_z \rangle$ in the short-time stage is much faster than that in the long-time
stage. It is also worth noting that a strong oscillation can exist in the
short-time stage, as shown in Fig.~\ref{fig2}(c). The evolution of $\langle S_z
\rangle$ can be fitted by a double exponential function
$f_2(t)=a_1e^{-t/\tau_{\rm{E_1}}}+a_2e^{-t/\tau_{\rm{E_2}}}+b$ [orange-red solid
curves in Fig.~\ref{fig2}(b)$\sim$(d)] with $a_1$, $a_2$, $b$, $\tau_{\rm{E_1}}$
and $\tau_{\rm{E_2}}$ being fitting parameters.

The behavior of $\langle S_z \rangle$ suggests that there are two main spin
relaxation processes in the system: the first process with the characteristic
rate $\tau^{-1}_{\rm{E_1}}$, which dominates in the short-time stage and the
second process with the characteristic rate $\tau^{-1}_{\rm{E_2}}$, which
dominates in the long-time stage. If the time evolution of $\langle S_z \rangle$
belongs to type (i), only one process dominates (corresponding to
$\tau^{-1}_{\rm{E_1}}$), which solely determines the effective SRRs. If the time
evolution of $\langle S_z \rangle$ belongs to type (ii), both processes can
contribute to the effective SRRs. The effective SRR
$\bar{\tau}^{-1}_{\rm{ \alpha}}$ is given by the average over the two rates
$\tau^{-1}_{\rm{E_1}}$ and $\tau^{-1}_{\rm{E_2}}$ with the weight determined by
the parameter $\alpha$. For small $\alpha$, the effective SRR
$\bar{\tau}^{-1}_{\rm{\alpha}}$ mainly describes the short time behavior and
hence has a large weight of $\tau^{-1}_{\rm{E_1}}$. While for large $\alpha$,
$\tau^{-1}_{\rm{E_2}}$ makes a large contribution to the
$\bar{\tau}^{-1}_{\rm{\alpha}}$.

To give a comprehensive description of both the short- and long-time
behaviors of the spin relaxation, we choose the effective SRR
$\bar{\tau}^{-1}_{\rm{\alpha}}$ with two typical values of
$\alpha$: $\alpha=1.5$ which better describes the short-time behavior
and $\alpha=e$ which better describes the long-time behavior.

\subsubsection{Electric field dependence}

The effective SRRs $\bar{\tau}^{-1}_{\rm{1.5}}$ and
$\bar{\tau}^{-1}_{\rm{e}}$ as function of electric field at magnetic field
$B=1$~T and temperature $T=0$~K are plotted in Fig.~\ref{fig2}(e). One can see
that both effective SRRs have similar electric field dependence with two main
features: one is the large drop at the anti-crossing point, the other is the
multiple sharp peaks indicated by the numbered arrows.

To understand these features, it is helpful to first identify the dominant spin
relaxation process at different electric fields. This can be done by comparing
the effective SRRs $\bar{\tau}^{-1}_{\rm{1.5}}$ and
$\bar{\tau}^{-1}_{\rm{e}}$ to the two rates $\tau^{-1}_{\rm{E_{1}}}$ and
$\tau^{-1}_{\rm{E_{2}}}$ in Fig.~\ref{fig2}(e). Far from the anti-crossing point
($E<0.102$~kV/cm and $E>0.19$~kV/cm), only one spin relaxation process
dominates. Both effective SRRs are determined by the corresponding rate
$\tau^{-1}_{\rm{E_{1}}}$. For the electric field $E$ in the vicinity of the
anti-crossing point [$E \in (0.102, 0.19)$~kV/cm], the second spin relaxation
process becomes important and can dominate the effective SRRs after the
anti-crossing point. The difference of the two effective SRRs
$\bar{\tau}^{-1}_{\rm{1.5}}$ and $\bar{\tau}^{-1}_{\rm{e}}$ at the
anti-crossing point is due to the fact that the rate $\tau^{-1}_{\rm{E_{1}}}$
has a smaller contribution to $\bar{\tau}^{-1}_{\rm{e}}$ than to
$\bar{\tau}^{-1}_{\rm{1.5}}$.

Having identified the dominant spin relaxation process, one finds that
the first feature, e.g., the large drop at the anti-crossing point, is
due to the relative importance of the two spin relaxation processes:
the second spin relaxation process becomes more and more important as
the electric field $E$ passes the anti-crossing point. The second
feature, e.g., the multiple sharp peaks, is solely induced by the spin
relaxation process corresponding to $\tau^{-1}_{\rm{E_{1}}}$.

\begin{figure}
  \begin{center}
    \includegraphics[width=7.5cm]{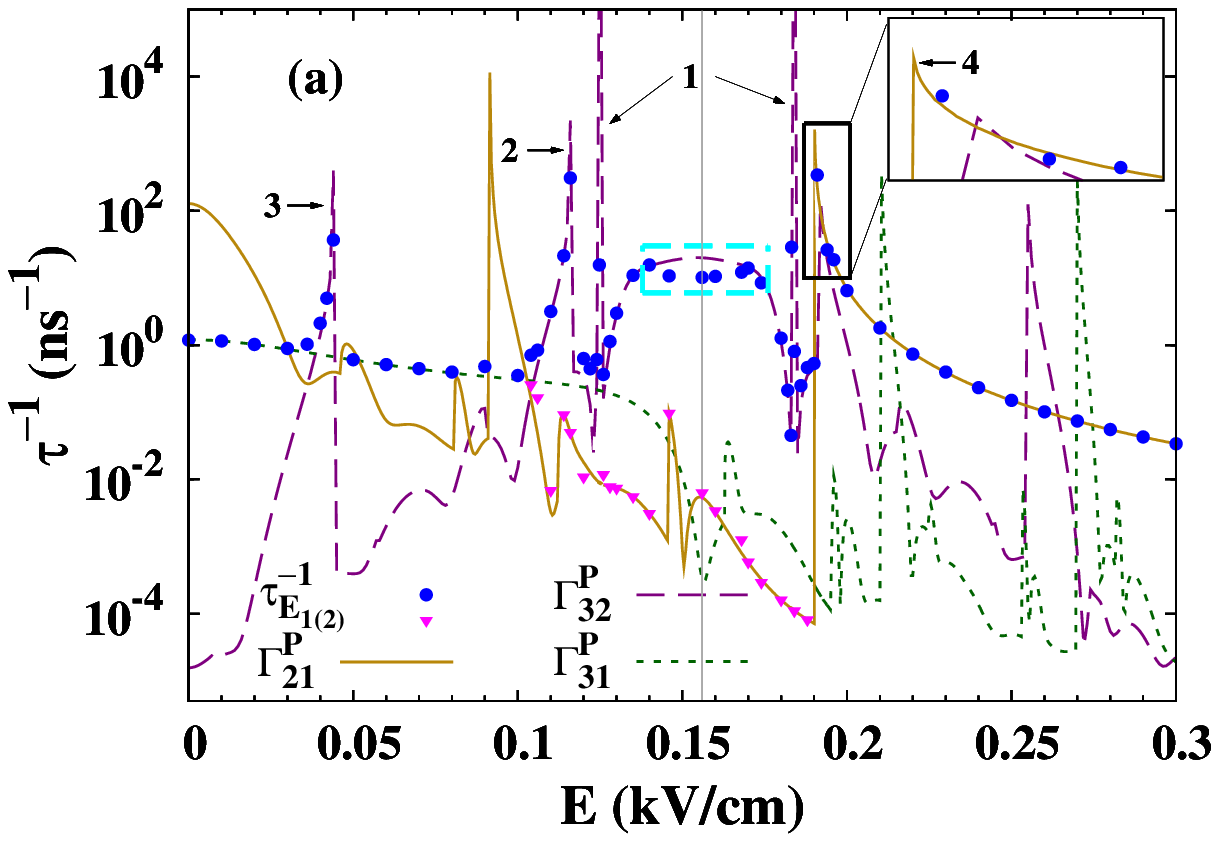}
    \includegraphics[width=7.5cm]{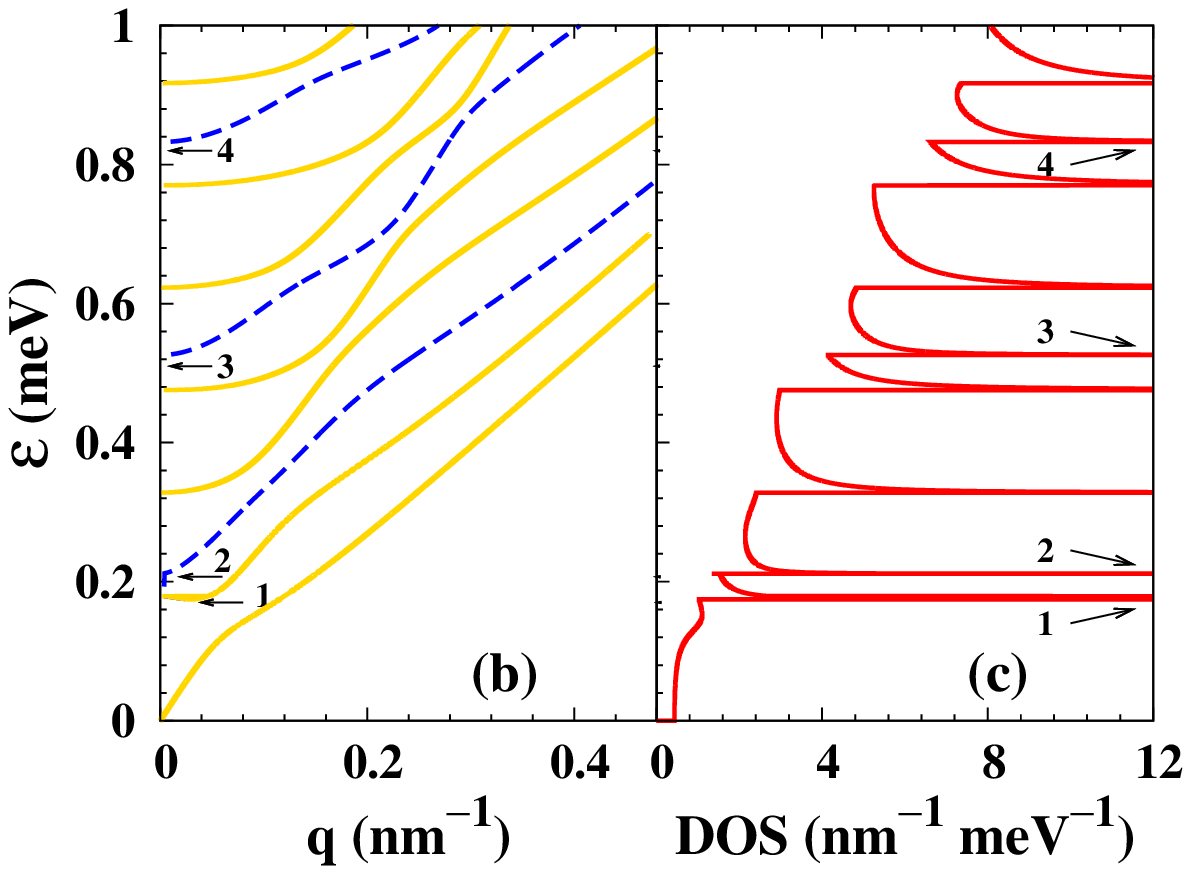}
  \end{center}
  \caption{(Color online) (a) Transition rates $\Gamma^{\rm P}_{\rm 21}$ (yellow
    solid curve), $\Gamma^{\rm P}_{\rm 32}$ (purple dashed curve) and
    $\Gamma^{\rm P}_{\rm 31}$ (green dotted curve) due to the piezoelectric
    coupling only, with $B=1$~T and $T=0$~K. Blue dots/Pink triangles: rate for
    the first/second spin relaxation process. Inset: the enlargement of the peak
    4. The cyan dashed box indicates the vicinity of the anti-crossing point and
    the gray vertical line indicates the position of the anti-crossing point,
    which are the same with those in Fig.~\ref{fig2}(b). (b) Spectrum of the
    confined phonons in the nanowire. The orange solid/blue dashed curves
    represent the axial/radial phonon modes. (c) DOS of the confined
    phonons. The arrows with numbers in (b) and (c) indicate the van Hove
    singularities which are responsible for the sharp peaks labelled with the
    same numbers in (a).}
  \label{fig3}
\end{figure}

The relative importance of the two processes can be understood by studying the
relevant energy levels. This can be done by comparing the transition rates
between the lowest three energy levels to the rates $\tau^{-1}_{\rm{E_{1}}}$ and
$\tau^{-1}_{\rm{E_{2}}}$ in Fig.~\ref{fig3}(a). The first spin relaxation
process corresponding to $\tau^{-1}_{\rm{E_{1}}}$ is determined by different
transitions at different electric fields. For $E < 0.19$~kV/cm, this process is
mainly determined by the transition $|3\rangle \to |1\rangle$ or $|3\rangle \to
|2\rangle$. The transition with larger transition rate dominates this
process. For $E > 0.19$~kV/cm, this process is determined by the transition
$|2\rangle \to |1\rangle$. The second spin relaxation process corresponding to
$\tau^{-1}_{\rm{E_{2}}}$, which only exists in the regime $E \in (0.102,
0.19)$~kV/cm, is determined by the transition $|2\rangle \to |1\rangle$. So the
dominant spin relaxation process changes from $|3\rangle \to |2\rangle$ to
$|2\rangle \to |1\rangle$ as the electric field $E$ passes the anti-crossing
point, which agrees with the spins of the states shown in
Fig.~\ref{fig1}(a). This explains the large drop at the anti-crossing
point. Note that in the calculation of the transition
rate, we consider only the piezoelectric coupling since it is dominant in our
system.\cite{cweber1}

Now let us turn to the second feature of the effective SRRs, e.g., the multiple
sharp peaks, which are determined by the first spin relaxation process
corresponding to $\tau^{-1}_{\rm{E_{1}}}$. These sharp peaks occur as the energy
splitting between the two states of the spin relaxation process matches the van
Hove singularities of the confined phonons, e.g., the denominator of the
transition rate $\Gamma^{\rm P}_{\ell_1 \ell_2}$ tends to zero.\cite{yyin}
According to Fig.~\ref{fig3}(a), one can see that the peaks 2, 3 and the double
peaks 1 are induced by the transition $|3\rangle \to |2\rangle$ whereas the peak
4 is induced by the transition $|2\rangle \to |1\rangle$. By comparing the
energy levels and corresponding spin configurations of the DQD shown in
Fig.~\ref{fig1}(a) to the spectrum and DOS of the confined phonons in
Fig.~\ref{fig3}(b) and (c), one can also identify the corresponding confined
phonon mode and van Hove singularity for each peak, which are indicated by the
arrows with the same numbers in Fig.~\ref{fig3}(b) and (c), respectively.

It should be emphasized that not every van Hove singularity can induce a sharp
peak. This is because the behavior of the nominator of the coefficient
$\Gamma^{\rm P}_{\ell_1 \ell_2}$, e.g., the form factor, is also crucial for the
peaks, which depends on the properties of the confined phonon modes and the
electron-phonon interactions. Zeros of the form factors can greatly suppress the
SRR. To induce a sharp peak, the suppression must be absent or weak at the
corresponding van Hove singularity. Two types of van Hove singularities satisfy
such condition in our system. The first type is the van Hove singularities with
$q=0$ for the radial phonon modes. The peaks (peaks 2, 3 and 4) exist since the
corresponding form factor of the piezoelectric coupling tends to zero slow
enough so that the suppression effect is weak. The second type is the van Hove
singularities with phonon wave vector $q \ne 0$. The corresponding peaks (the
double peaks 1) are divergent due to the absence of zeros of the form factors at
$q \ne 0$.\cite{yyin} Both the deformation potential and piezoelectric couplings
can induce such peaks.

These peaks can be used for electric spin relaxation manipulation. For example,
around peak 3, by changing the electric field within $0.01$~kV/cm, the effective
SRR can change up to 2 orders of magnitude. While for the electric field a
little far away from peak 3, the effective SRRs become rather insensitive to the
electric field. Thus it can be used as electric on-and-off spin switches. It is
worth comparing our spin relaxation manipulation scheme to the one based on GaAs
DQDs in quantum wells where bulk phonons dominate discussed by Wang and Wu in
Ref.~\onlinecite{yywa}. There, the spin relaxation manipulation relies on the
fact that, for the vertical DQDs considered in their scheme, the electron-phonon
scattering between the bonding and anti-bonding orbital states of the DQDs has a
very large contribution to the SRR for small electric field. The SRR can be
greatly suppressed by breaking these states by strong electric field, resulting
in the spin relaxation manipulation. As the bonding/anti-bonding state plays the
central role in the quantum-well based DQD scheme, the high energy anti-bonding
state must be occupied or serve as the virtual state for the hyperfine-mediated
spin relaxation process. So their scheme works with high electric field (to
break the bonding/anti-bonding states) at high temperature (to make the high
energy anti-bonding state occupied) or weak magnetic field (to ensure the
hyperfine-mediated spin relaxation is not suppressed). In our scheme, the spin
relaxation manipulation relies mainly on the quasi-one-dimensional properties of
the confined phonons, which does not suffer from such restrictions. It can also
work at low temperature, strong magnetic field or weak electric field. Moreover,
the spin relaxation manipulation can be accomplished at different electric
fields, which can be tuned by the nanowire radius $R$. Thus our scheme is more
sensitive and flexible than the scheme in quantum-well-based DQDs.

In the above discussion, facilitated with the transition rates, we have
explained the two main features of the effective SRR from the equation-of-motion
approach. It should be emphasized that in such explanation based on the
transition rate, only the contribution of the diagonal elements of the density
matrix to the spin relaxation is considered. This is valid when the electric
field is far away from the anti-crossing point. However, for the electric field
sufficiently close to the anti-crossing point, the contribution of the
off-diagonal elements becomes important, which can have a pronounced effect on
the spin relaxation. The spin relaxation in this case cannot be explained by the
transition rate. This can be seen in the cyan dashed box in Fig.~\ref{fig3}(a)
where $\Gamma^{\rm P}_{\rm 32}$ overestimates the rate $\tau^{-1}_{\rm{E_{1}}}$
induced by the transition $|3\rangle \to |2\rangle$.

In fact, the rate $\tau^{-1}_{\rm{E_{1}}}$ at the anti-crossing point
$E=0.156$~kV/cm is equal to $\Gamma^{\rm P}_{\rm 32}/2$, which is just the
relaxation rate of the off-diagonal elements between the states $|3\rangle$ and
$|2\rangle$. This suggests that the off-diagonal elements dominate the spin
relaxation at this point. This can be seen more clearly from the temperature
dependence of the effective SRR in the next section.

\subsubsection{Temperature dependence}

\begin{figure}
  \begin{center}
    \includegraphics[width=7.5cm]{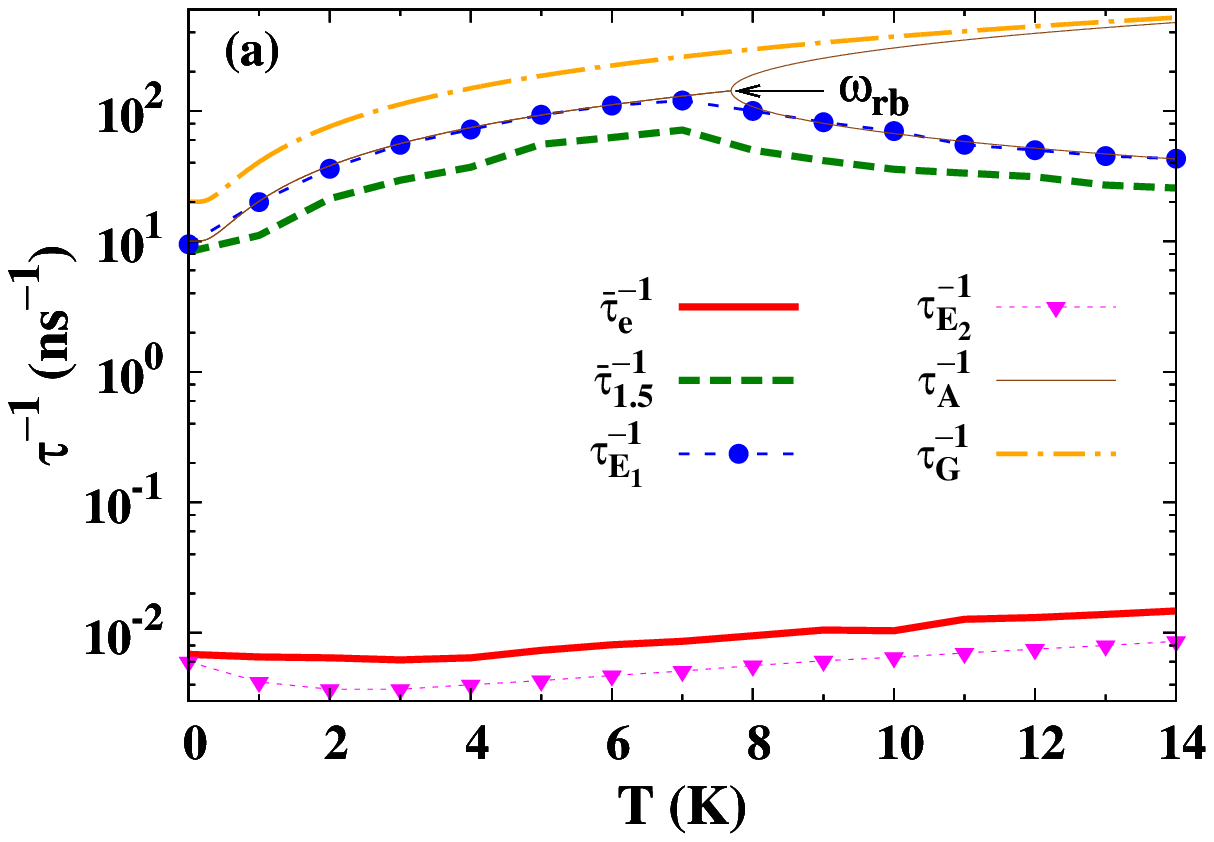}
    \includegraphics[width=7.5cm]{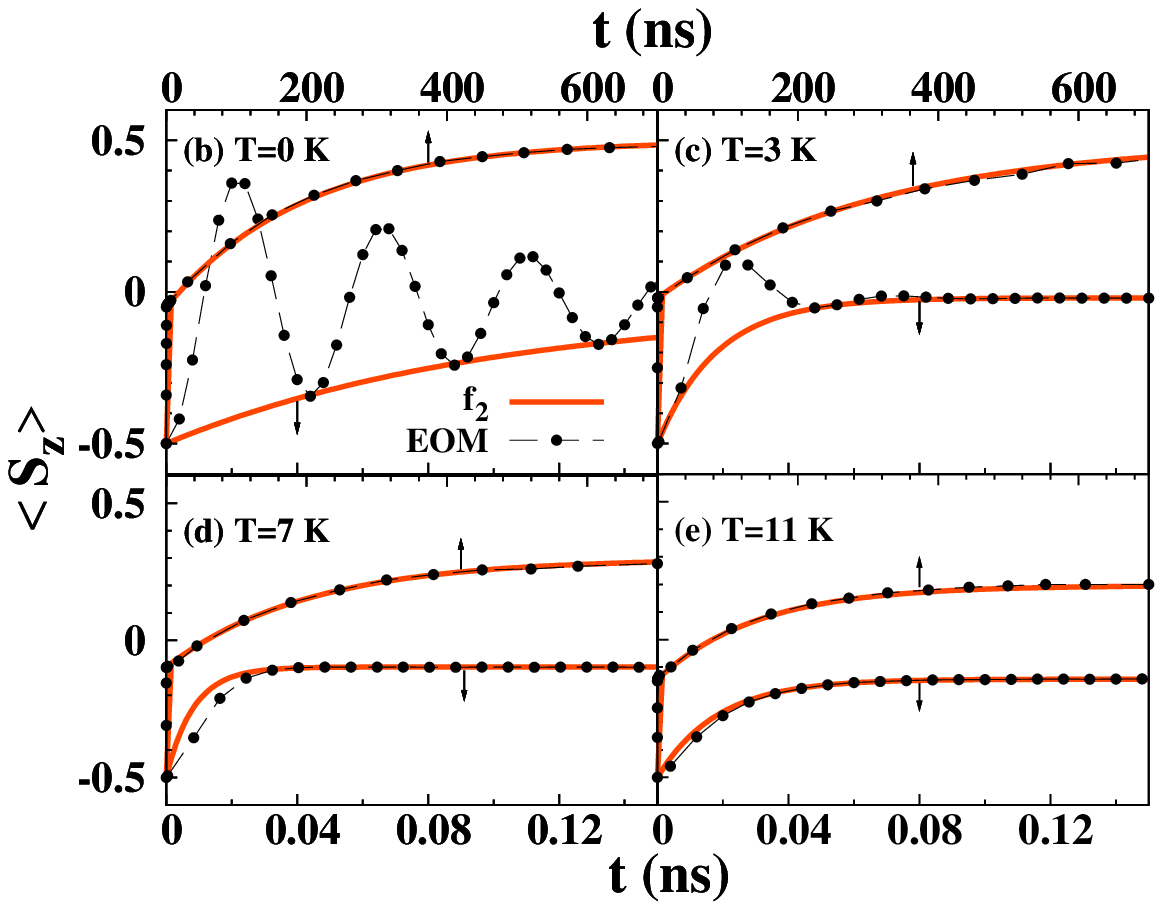}
  \end{center}
  \caption{(Color online) (a) SRRs as function of the temperature $T$ with
    $E=0.156$~kV/cm and $B=1$~T. Curves (dots) with the same color and type
    represent the same quantities in Fig.~\ref{fig2}(e). Brown solid thin
    curve: the relaxation rate from Eq.~(\ref{eq20}). Yellow dashed-dotted curve:
    the SRR from the Fermi-golden-rule approach. See text for details. (b)
    $\sim$ (e) Time evolution of $\langle S_z\rangle$ in both the short- (lower
    scale) and long-time (upper scale) regimes with $E=0.156$~kV/cm and $B=1$~T
    at temperature $T=0$, $3$, $7$ and $11$~K, respectively. Curves with the
    same color and type represent the same quantities in Fig.~\ref{fig2}}
  \label{fig4}
\end{figure}

Now we discuss the temperature dependence of the effective SRRs at the
anti-crossing point with $E=0.156$~kV/cm and $B=1$~T, shown in
Fig.~\ref{fig4}(a). In the temperature regime we consider here, the contribution
of the high energy levels can still be neglected, so only the lowest three
energy levels are important. The most prominent feature of the effective SRRs is
the existence of the smooth peak for $\bar\tau^{-1}_{\rm{1.5}}$ at $T=7$~K,
while $\bar\tau^{-1}_{\rm{e}}$ is monotonically increasing. Comparing to the
two rates $\tau^{-1}_{\rm{E_1}}$ and $\tau^{-1}_{\rm{E_2}}$, one finds that
$\bar\tau^{-1}_{\rm{1.5}}$ is mainly determined by the spin relaxation
process corresponding to $\tau^{-1}_{\rm{E_1}}$, which is due to the transition
$|3\rangle \to |2\rangle$ at the anti-crossing point. While
$\bar\tau^{-1}_{\rm{E_{e}}}$ is mainly determined by $\tau^{-1}_{\rm{E_2}}$
which is induced by the transition $|2\rangle \to |1\rangle$. The peak is mainly
due to the transition $|3\rangle \to |2\rangle$, so the peak in
$\bar\tau^{-1}_{\rm{1.5}}$ is absent in $\bar\tau^{-1}_{\rm{e}}$. Note
that the behaviors of the time evolution of $\langle S_z \rangle$ from the
equation-of-motion approach with the temperature before and after the peak are
different, which can be seen in Fig.~\ref{fig4}(b)$\sim$(e). Although both can
be fitted by the double exponential function, $\langle S_z \rangle$ exhibits an
oscillation in the first spin relaxation process before the peak
[Fig.~\ref{fig4}(b)(c)] while the oscillation vanishes after the peak
[Fig.~\ref{fig4}(d)(e)].

The above-mentioned behavior can be understood if we assume the
off-diagonal elements of the density matrix between the two states
$|2\rangle$ and $|3\rangle$ dominate the spin relaxation. As far as
the two states are concerned, the time evolution of $\langle S_z
\rangle$ due to the off-diagonal elements follows an oscillation
decay: $\langle S_z(t) \rangle = A \sin(\Omega t) e^{-t/T_2} + B$,
where $\Omega = \sqrt{\omega^2_{\rm rb} - T^{-2}_2}$ with $\omega_{\rm
  rb}$ representing the Rabi frequency and $T^{-1}_2=\Gamma^{\rm
  P}_{\rm 32}/2$ being the relaxation rate. The parameters $A$ and $B$
are constants. As the temperature increases, the oscillation becomes
slower due to the increase of the transition rate $\Gamma^{\rm
  P}_{\rm 32}$ and vanishes when $T^{-1}_2 > \omega_{\rm rb}$,
contributing another exponent decay to $\langle S_z \rangle$. This
explains the vanishing of the oscillation in $\langle S_z \rangle$
from the equation-of-motion approach at high temperature.

According to the previous discussion, one can also give an analytical estimation
for the relaxation rate induced by the off-diagonal elements of the density
matrix between $|3\rangle$ and $|2\rangle$:
\begin{equation}
  \tau^{-1}_{\rm A}= T^{-1}_2 \pm \theta( T^{-1}_2- \omega_{\rm rb})
  \tilde{\Omega},
  \label{eq20}
\end{equation}
where $\tilde{\Omega} = \sqrt{T^{-2}_2 - \omega^2_{\rm rb}}$. We plot
$\tau^{-1}_{\rm A}$ in Fig.~\ref{fig4}(a), which agrees with
$\tau^{-1}_{\rm{E_1}}$ very well. Note that the upper branch of $\tau^{-1}_{\rm
  A}$ [corresponding to ``+'' in Eq.~(\ref{eq20})] cannot be obtained from the
equation-of-motion approach since its contribution to the spin relaxation is
very small thus is difficult to be fitted numerically.

It is noted that this peak is difficult to obtain for the quantum-well based QDs
where bulk phonons dominate. This is mainly due to the fact that the transition
rate is suppressed by the small DOS of the bulk phonons when the energy
splitting between the two states becomes small,\cite{golovach} making it
difficult to fulfill the condition $T^{-1}_2 > \omega_{\rm rb}$.  Whereas the
DOS of the confined phonons is almost constant away from the van Hove
singularities, it is easy to have the condition $T^{-1}_2 > \omega_{\rm rb}$
satisfied.

It is pointed out that the large contribution of the off-diagonal elements makes
the widely-used Fermi-golden-rule approach,\cite{jlch,jhji,psta,dvbu,cfde,vaab}
which calculates the SRR by the formula $\tau^{-1}_{\rm G}=\sum_{\eta={\rm
    D},{\rm P}}\sum_{\ell_1\ell_2} F_{\ell_1}\Gamma_{\ell_1\ell_2}^{\eta}$,
inapplicable. As an example, we plot the temperature dependence of the SRR
$\tau^{-1}_{\rm G}$ from the Fermi-golden-rule approach in Fig.~\ref{fig4}(a)
with yellow dashed-dotted curve. It exhibits an monotonic increase without the
temperature peak. Moreover, it overestimates the SRR in the whole temperature
regime. This suggests that for the system with strong spin mixing, the
Fermi-golden-rule approach is inadequate. To calculate the SRR properly, one has
to apply the equation-of-motion approach.

\section{SUMMARY}

In summary, we have investigated the effective SRRs induced by confined phonons
in the nanowire-based DQDs. We first study the energies and spins of the lowest
three states and find that both can be efficiently tuned by the electric
field. An anti-crossing exists between the first and second excited states,
which is due to the coaction of the electric field, the magnetic field and the
SOC. Then we study the electric field dependence of the effective SRR within the
equation-of-motion approach and find that multiple sharp peaks exist in the
effective SRR, which can be used as the electric on-and-off spin switches. The
confined phonons are found to be crucial for these peaks. The
quasi-one-dimensional confined phonons exhibit extremely strong DOS at the van
Hove singularities, resulting in the multiple sharp peaks as the energy
splittings between states with different spins match certain van Hove
singularities. Both the van Hove singularities with $q\ne 0$ and the ones with
$q=0$ of radial confined phonon modes can induce such peaks, whilst the
piezoelectric coupling is found to be crucial for the latter. The manipulation
of the spin relaxation based on these multiple sharp peaks offers on-and-off
spin switches, which is more sensitive and flexible than the one based on DQDs
in quantum wells. We also find that the contribution of the off-diagonal
elements of the density matrix can dominate the spin relaxation at the
anti-crossing point due to the large spin mixing, resulting in a smooth peak in
the temperature dependence of the effective SRR. This makes the widely-used
Fermi-golden-rule approach inadequate in the vicinity of the anti-crossing
point. These results show a great potential of the nanowire-based DQDs in the
spin-based quantum information processing and spintronic devices.

\begin{acknowledgments}
  This work was supported by the Natural Science Foundation of China
  under Grant No.\ 10725417. One of the authors (YY) was also
  partially supported by the China Postdoctoral Science Foundation.
\end{acknowledgments}

\end{document}